\documentclass[aps,prl,twocolumn, 10pt, showpacs]{revtex4}

\usepackage{graphicx}% Include figure files
\usepackage{dcolumn}% Align table columns on decimal point
\usepackage{bm}% bold math
\usepackage{amsmath}
\usepackage{amssymb}
\usepackage{footnpag}

\begin{document}
%\thispagestyle{empty}
%%%%%%%%%%%%%%%%%%%%%%%%%%%%%%%
\title{Percolation and Colossal Magnetoresistance in Eu-based Hexaborides}

\author{G.A. Wigger}
\author{C. Beeli}
\author{E. Felder}
\author{H. R. Ott}

\affiliation{Laboratorium f\"{u}r Festk\"{o}rperphysik,
ETH-H\"{o}nggerberg, CH-8093 Z\"{u}rich, Switzerland }

\author{A.D. Bianchi}
   \altaffiliation{Present address: Hochfeldlabor Dresden Forschungszentrum Rossendorf, Postfach 51 01 19, 01314, Dresden Germany}
\author{Z. Fisk}

\affiliation{National High Magnetic Field Laboratory, Florida
State University, Tallahassee, Florida 32306}

\date{\today}

\begin{abstract}
Upon substituting Ca for Eu in the local-moment ferromagnet
EuB$_6$, the Curie temperature $T_C$ decreases substantially with
increasing dilution of the magnetic sublattice and is completely
suppressed for $x$ $\leq$ 0.3. The Ca substitution leads to
significant changes of the electronic properties across the
Eu$_x$Ca$_{1-x}$B$_6$ series. Electron microscopy data for $x$
$\approx$ 0.27 indicate a phase separation into Eu- and Ca-rich
clusters of 5 to 10 nm diameter, leading to percolation-type
phenomena in the electrical transport properties. The related
critical concentration $x_p$ is approximately 0.3. For $x$
$\approx$ 0.27, we observe colossal negative magnetoresistance
effects at low temperatures, similar in magnitude as those
reported for manganese oxides.
\end{abstract}
\maketitle

Strong variations of the electrical resistivity $\rho$ by external
magnetic fields may be achieved either in specially tailored
thin-film heterostructures, leading to so called giant
magnetoresistance \cite{fert1}, or in manganese oxides which
exhibit metal/insulator transitions and concomitant magnetic
ordering phenomena, leading to colossal magnetoresistance effects
\cite{jin}. Giant magnetoresistance is related to the
spin-dependent scattering of conduction electrons between
magnetically ordered layers which is strongly affected by external
fields \cite{fert2}. Colossal magnetoresistance is traced back to
a strong coupling of the electronic subsystem to the lattice,
leading to polaron formation \cite{batlogg}. The onset of
ferromagnetic order, favored by applying external fields, induces
a transition to a metallic, i.e., low resistive state.

Large magnetoresistance effects are also known to occur in some
Eu-based chalcogenides \cite{kasuya} and in EuB$_6$
\cite{bachmann}. Also here, ferromagnetism favors metallicity,
both by enhancing the number of itinerant charge carriers and by
reducing the spin disorder scattering \cite{leo}. Again,
increasing magnetic fields help to enhance the electrical
conductivity of the respective materials.

In this work, we present and discuss a set of electrical transport
data for materials of the series Eu$_{x}$Ca$_{1-x}$B$_6$ and
demonstrate that they are strongly influenced by a phase
separation between nanoscopically small Eu- and Ca-rich regions.
This phenomenon leads to percolation effects in both the onset of
magnetic order and the electronic transport. For $x$ close to
$x_p$ = 0.31, the three dimensional site percolation limit for a
simple cubic lattice \cite{kirkpatrick1, efros, efros2}, we
observe extremely large negative magnetoresistance effects.

\begin{figure}
  \centering
  \includegraphics[width=\linewidth]{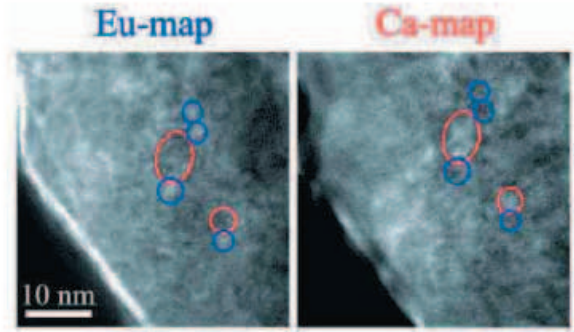}
  \caption{(Color in online edition) EFTEM images for $x$ $\sim$ 0.27. In the left (right) panel, the energy window
  was chosen such that the electrons scatter inelastically off the Eu- (Ca-)
  ions. The red (blue) rings mark regions rich in Ca (Eu). Bright
  (dark) regions indicate a high (low) intensity of scattered
  electrons}
 \label{eucamap}
\end{figure}

\begin{figure}
  \centering
  \includegraphics[width=\linewidth]{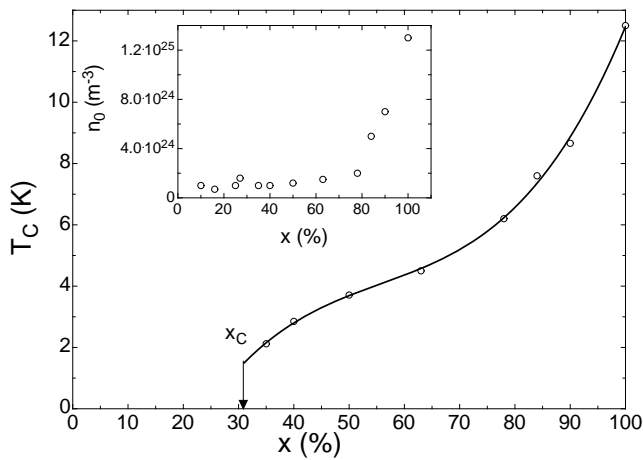}
  \caption{$T_C$ obtained from magnetization, specific heat and resistivity data.
  The solid line is to guide the eye.
  The critical value $x_C$ represents the limit for the site percolation in a cubic 3D lattice. The inset displays
  the itinerant electron density $n_0(x)$, which is approximately constant between 30 and 70 K for all values of
  $x$.}
 \label{tceuca}
\end{figure}

Single-crystalline samples of Eu$_x$Ca$_{1-x}$B$_6$ were grown in
a flux of 99.999$\%$ pure Al, using the starting elements Eu, Ca
and B with purities of 99.95$\%$, 99.987$\%$ and 99.99$\%$,
respectively. Measurements of the magnetization were made with a
commercial SQUID magnetometer. In order to avoid the influence of
possible magnetic impurities at the surface, the samples were
etched for a short time in strongly diluted nitric acid. High
resolution transmission electron microscopy (HRTEM) investigations
were made with a FEI Technai F30 instrument, equipped with a field
emission electron gun and an energy filter from Gatan Inc. The
filter employs an electron energy loss spectrometer (EELS) to
select a particular energy window for the inelastically scattered
electrons to form the image. The chemical maps shown in
Fig.(\ref{eucamap}) were recorded with the so-called three-window
technique. Selected area electron diffraction patterns and HRTEM
images, which are both not shown because of space restrictions,
reveal the high perfection of the atomic arrangements in the
crystals, also for low values of $x$. However, energy-filtered TEM
images reveal a phase separation resulting in Eu- and Ca-rich
regions. An example is shown in Fig.(\ref{eucamap}) for $x$ =
0.27. Two patches relatively rich in Ca are marked by red rings,
several other patches relatively rich in Eu are emphasized by blue
rings. The comparison of the two maps reveals that a relative
richness in one element is correlated with a relative deficiency
in the other. Analogue Boron maps confirm a constant distribution
of the anion element. The typical diameter of the clusters is
between 5 and 10 nm. In view of the different concentrations of
charge carriers in the different regions (see inset of
Fig.(\ref{tceuca})), the maps shown in Fig.(\ref{eucamap}) may
also be taken as evidence for an electronic phase separation and
hence the formation of regions with different electrical
conductivities.

The resistivities and the Hall voltages were measured using a
standard low-frequency a.c. four-contact technique in longitudinal
and transversal configuration, respectively. The covered
temperature and magnetic field ranges were between 0.35 and 300 K,
and 0 and 7 Tesla, respectively.

In Fig.(\ref{tceuca}) we present the $x$ dependence of the Curie
temperature $T_C(x)$ as established by measurements of the
specific heat $C_p(T)$ and the electrical resistivity $\rho(T)$ in
zero magnetic field, as well as by Arrott-plot analyses of the
magnetization $M(T,H)$. $T_C$ decreases monotonously but non
linearly with decreasing $x$ and vanishes at $x$ $\sim$ 0.3. The
non-linear decrease of $T_C(x)$ most likely reflects the formation
of Eu- and Ca-rich regions, whereby $x$ merely represents an
average concentration of Eu ions. For $x$ $\leq$ 0.3, i.e., below
the percolation limit, the ground state exhibits spin-glass type
features \cite{future}.

In the following we argue that the phase separation is also
reflected in the electronic transport properties across the
series. As described in detail before
\cite{mandrus,wiggermonnier}, the total electrical resistivity
$\rho$ of these materials is due to different contributions. In
our case we consider

\begin{equation}\label{matthiessens}
  \rho = \rho_{sd} + \rho_{e-ph} + \rho_d    \ \ \ ,
\end{equation}

where $\rho_{sd}$, $\rho_{e-ph}$ and $\rho_d$ represent the
spin-disorder, the electron-phonon and the defect scattering,
respectively.

Because in these hexaborides, the concentration of conduction
electrons is temperature dependent, all three contributions to
$\rho(T)$ vary with temperature, in particular also $\rho_d$. It
is $\rho_d$ which is mainly affected by the phase separation and
therefore is expected to reflect the influence of percolation
effects most significantly.

\begin{figure}
  \centering
  \includegraphics[width=\linewidth]{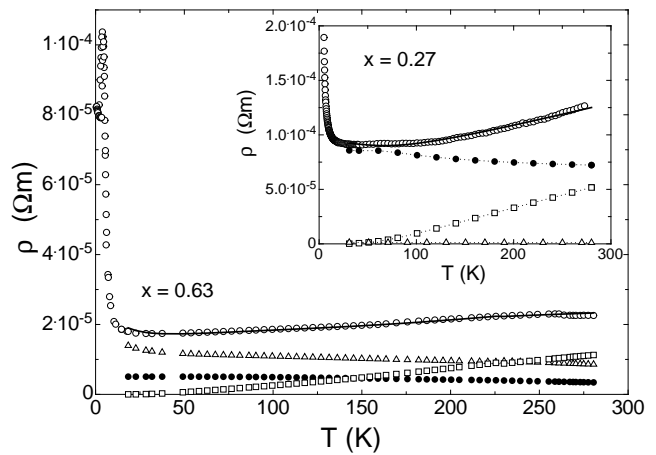}
  \caption{Separation of $\rho(T)$ (open circles) for material with $x$ =
  0.63 into spin-disorder (triangles), electron-phonon (squares) and
  defect-induced resistivity (full circles). The solid line is the result of our calculation described in the text. The inset
  shows the same for $x$ = 0.27. Here, the dotted lines are to guide the eye.}
\label{hightemps}
\end{figure}

The spin disorder scattering is taken into account by
\cite{taylordarby, haas}

\begin{equation}\label{haaseq}
  \rho_{sd} = \frac{1}{4\pi^{1/3}}\frac{N
  m^*}{\hbar^3e^2}\left(\frac{J}{N}\right)^2xS(xS+1)\frac{T}{T-T_C}n^{-2/3}
  \ \ \ ,
\end{equation}

with $J$ as the effective exchange interaction parameter, $N$ as
the number of cations per unit volume, $m^{*}$ as the effective
mass and $n$ as the concentration of the conduction electrons. For
the scattering of electrons on phonons, we used the same model
approach that was successful in previously published analyses of
the electrical resistivities of LaB$_6$ \cite{mandrus} and EuB$_6$
\cite{wiggermonnier}, where the phonon spectrum was approximated
by a Debye type spectrum and two superimposed Einstein modes. The
latter modes model the motion of Eu and Ca ions inside the
respective Boron cages. The Debye temperature was taken as 1160 K,
as previously reported \cite{mandrus, wiggermonnier}. The best
fits were obtained with Einstein temperatures of 168 K and 373 K
for the Eu and the Ca mode, respectively. The activation of mobile
charge carriers out of defect states, implied by results of Hall
voltage measurements \cite{future, wiggermonnier}, was considered
by setting

\begin{equation}\label{activation}
  n(x,T) = n_0(x) + \Delta n(x)\cdot e^{-E_{ex}(x)/k_BT} \ \ \ ,
\end{equation}

where $n_0(x)$ represents the constant mobile-carrier densities in
the range 30 K $\leq$ $T$ $\leq$ 70 K, shown in the inset of
Fig.(\ref{tceuca}). The excitation energy $E_{ex}(x)$ is the
energy of defect states with respect to the Fermi energy. The
relaxation rate for the defect scattering, assumed to be
temperature independent, was evaluated from $\rho_d$ = $\rho$ -
$\rho_{sd}$ - $\rho_{e-ph}$ and $n_0(x)$.

\begin{figure}
  \centering
  \includegraphics[width=\linewidth]{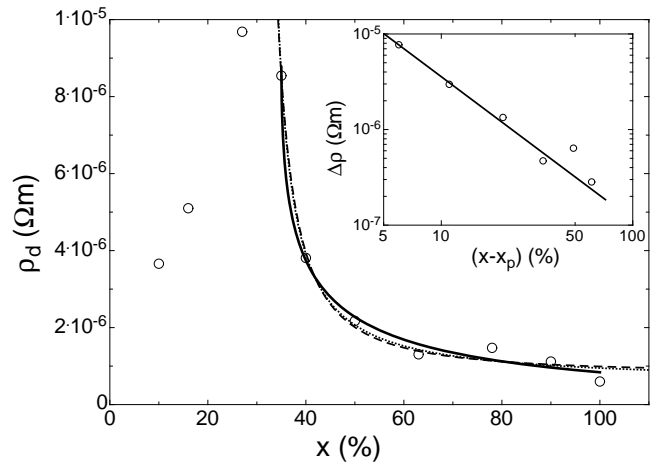}
  \caption{$\rho_d(x)$. The dotted and the broken line describe fits
  according eq. \ref{powerrho} with $t$ = 1.2 and 1.5, respectively. The solid line is
  a solution to eq. \ref{mclachlansigma}. The inset shows
  $\Delta\rho$ = $\rho_d - \rho_{\infty}$
  vs. $x-x_p$ on a log-log scale. This verifies the power law with
  exponent 1.5.}
 \label{residualrho}
\end{figure}

The temperature dependences of the individual contributions,
adding up to the total $\rho(T)$ of material with $x$ = 0.63 and
0.27, is shown in Fig.(\ref{hightemps}). The calculated solid
lines are in remarkably good agreement with experiment. In order
to emphasize the effect of percolation, $\rho_d(x)$ was scaled to
a charge carrier concentration of 1.3$\cdot$10$^{25}$ m$^{-3}$,
the value of $n_0(x=1)$. From fits to the experimental data for
$\rho(T)$ for other values of $x$, the scaled values of
$\rho_d(x)$ were evaluated and plotted in the main frame of
Fig.(\ref{residualrho}). The usual Nordheim-Kurnakov relation
\cite{ziman} for a two component system

\begin{equation}\label{nordheimkurnakov}
  \rho_d(x) \sim x(1-x)
\end{equation}

is not even qualitatively obeyed. Instead a tendency to divergence
at $x$ $\sim$ 0.33, close to the above mentioned site percolation
limit for a simple cubic system of $x_p$ = 0.31 is observed. A
percolation dominated resistivity is expected to be described by

\begin{equation}\label{powerrho}
  \rho_d = \rho_d'(x-x_p)^{-t} + \rho_{\infty} \ \ \ ,
\end{equation}

with 1.5 $\leq$ $t$ $\leq$ 1.6 \cite{kirkpatrick2, shklovskii,
dunn}. In our case, $\rho_{\infty}$ $\approx$ 4.2$\cdot$10$^{-7}$
$\Omega$m. A first attempt to fit the data for $x$ $>$ 0.35 and
assuming $x_p$ = 0.31, resulted in $t$ = 1.12 $\pm$ 0.05,
distinctly smaller than expected. The corresponding fit is
represented by the dotted line in Fig.(\ref{residualrho}). A
forced fit with $t$ = 1.5, shown as the broken line in
Fig.(\ref{residualrho}), results in $x_p$ = 0.29. Although no
perfect agreement between theoretical expectations and experiment
is achieved in this way, the qualitative behaviour of $\rho_d(x)$
strongly suggests that percolation plays a dominant role. This is
supported by the inset of Fig.(\ref{residualrho}), where $\rho_d$
- $\rho_{\infty}$ is plotted versus ($x$ - $x_p$) on double
logarithmic scales. The solid line represents a power law
variation with $t$ = 1.5.

Considering that both the Eu-, as well as the Ca-rich regions are
conducting, it seems reasonable to inspect whether a
phenomenological interpretation of the effective media theory due
to Bruggeman \cite{bruggeman}, treating two or more constituents
with different conductivities $\sigma_i$ and different volume
fractions $f_i$, is more appropriate for our case. This approach,
due to McLachlan \cite{mclachlan1}, yielded very good agreement
with experimental results by interpolating between the symmetric
and the antisymmetric solution of the Bruggeman model. The scheme
retains Bruggeman's solution in the limiting cases with the
correct values for $x_p$ and $t$. For our purpose we consider two
components; Eu-rich regions with a volume fraction $f$ and a high
conductivity $\sigma_h$, and Ca-rich regions with a volume
fraction $(1-f)$ and a low conductivity $\sigma_l$. Volume
fractions can be converted to atomic fractions by using

\begin{equation}\label{mclachlanfraction}
  f = \frac{R\cdot x}{R\cdot x + (1-x)}  \ \ \ ,
\end{equation}

where $x$ is the atomic bond or site fraction and $R$ is the
effective ratio of the two volumes. In our case, volume effects
are negligible because the lattice parameter changes only by about
1$\%$ across the entire series and we may set $f$ $=$ $x$ and
$f_c$ $=$ $x_p$. The conductivity of the effective medium is found
by solving \cite{mclachlan1}

\begin{equation}\label{mclachlansigma}
  \frac{f(\Sigma_h-\Sigma_m)}{\Sigma_h+\frac{f_c}{1-f_c}\Sigma_m}+\frac{(1-f)(\Sigma_l-\Sigma_m)}{\Sigma_l+\frac{f_c}{1-f_c}\Sigma_m}
  = 0 \ \ \ ,
\end{equation}

where $\Sigma_m$ is the scaled conductivity of the effective
medium and $\Sigma_i$ = $\sigma_i^t$ = $\rho_i^{-t}$ for $i$ =
$h$, $l$ and $m$. Based on this model we attempted to fit the
$\rho_d$ data by assuming $t$ = 1.5. The best result, represented
by the solid line in Fig.(\ref{residualrho}), yielded $\rho_h$ =
8.7$\cdot$10$^{-7}$ $\Omega$m and $\rho_l$ = 2.6$\cdot$10$^{-6}$
$\Omega$m, respectively. These values are close to $\rho_d(x=1)$
and, particularly rewarding, $\rho_d(x=0)$, respectively. The
percolation limit turns out to be $x_p$ $\sim$ 0.33, slightly
higher than the mentioned site percolation limit.

\begin{figure}
  \centering
  \includegraphics[width=\linewidth]{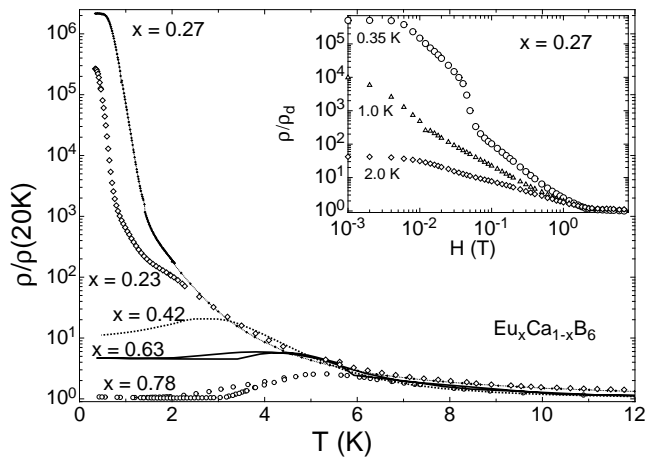}
  \caption{$\rho/\rho(20K)$ for $x$ = 0.23, 0.27, 0.42, 0.63, 0.78. The inset shows
  $\rho(H)/\rho_d$ for $x$ = 0.27 at 0.35, 1.0 and 2.0 K.}
 \label{resatxp}
\end{figure}

For material with $x$ close to but below $x_p$, we observe
remarkable magnetoresistance effects at very low temperatures. In
Fig.(\ref{resatxp}), $\rho(T)$ in zero external magnetic field is
shown for specimens with different values of $x$. The materials
with $x$ = 0.27 and 0.23 do not exhibit long range magnetic order
at low temperatures. Instead, spin-freezing type phenomena as
evidenced by the results of specific heat and a.c. susceptibility
measurements \cite{future} were observed. It is remarkable, that
$\rho(T)$ continues to increase dramatically below the
spin-freezing transitions at about 2 K. The inset of
Fig.(\ref{resatxp}) demonstrates that the strong enhancement of
$\rho$ at low temperatures can be eliminated by magnetic fields of
moderate strength. A more detailed discussion of these $\rho(H)$
data will be published elsewhere. The $\rho(T,0)$ curves are
qualitatively similar to those that were previously reported for
manganese oxides of the type (La$_{5/8-y}$Pr$_y$)Ca$_{3/8}$MnO$_3$
\cite{uehara}. It was claimed that for these materials, intrinsic
inhomogeneities due to phase separation are responsible for
colossal magnetoresistance effects. A theoretical interpretation
of those data, calculating the inverse conductivity for 3D
percolative systems \cite{mayr}, resulted in $\rho(T,0)$ curves of
qualitatively the same shape as those shown in
Fig.(\ref{resatxp}).

We intended to demonstrate that electronic phase separation,
related percolation and colossal magnetoresistance phenomena,
previously reported for selected oxide compounds \cite{uehara},
may also be found in systems as simple as the cubic hexaborides.
Although we cannot yet provide a convincing link between the
different phenomena in our case, ours and previous results
indicate a favorable route for future research efforts in relation
with colossal magnetoresistance effects.

Stimulating discussions with R. Monnier are gratefully
acknowledged. This work has benefitted from partial financial
support of the Schweizerische Nationalfonds zur F\"{o}rderung der
wissenschaftlichen Forschung and the US-NSF grant DMR-0203214.

\end{document}